\documentclass[fleqn,twoside]{article}
\usepackage{espcrc2}

\usepackage{graphicx}
\usepackage[figuresright]{rotating}

\newcommand{\AmS}{{\protect\the\textfont2
  A\kern-.1667em\lower.5ex\hbox{M}\kern-.125emS}}

\def\bar{\overline}
\def\phi{\varphi}

\newcommand{\beq}{\begin{equation}}
\newcommand{\eeq}{\end{equation}}
\newcommand{\bea}{\begin{eqnarray}}
\newcommand{\eea}{\end{eqnarray}}
\newcommand{\be}{\begin{displaymath}}
\newcommand{\ee}{\end{displaymath}}
\newcommand{\bc}{\begin{center}}
\newcommand{\ec}{\end{center}}
\newcommand{\bt}{\begin{tabbing}}
\newcommand{\et}{\end{tabbing}}

\newcommand{\reu}{{\nu_e\rightarrow\nu_\mu}}
\newcommand{\reub}{{\bar{\nu}_e\rightarrow\bar{\nu}_\mu}}
\newcommand{\ruu}{{\nu_\mu\rightarrow\nu_\mu}}
\newcommand{\ruub}{{\bar{\nu}_\mu\rightarrow\bar{\nu}_\mu}}
\newcommand{\rue}{{\nu_\mu\rightarrow\nu_e}}

\newcommand{\dm}[1]{{\Delta m^2_{#1}}}

\newcommand{\ie}{{\it i.e.}}

\newcommand{\eg}{{\it e.g.}}

\title{
\vspace*{-1.5cm}
\begin{flushright}
TUM-HEP-484/02\\
\end{flushright}
\vspace*{0.1cm}
\bf The Physics Potential of Future Long Baseline
Neutrino Oscillation Experiments$^*$}

\author{M.~Lindner
\address[MCSD]{Physik--Department, Technische Universit\"at M\"unchen,
James--Franck--Strasse, D--85748 Garching, Germany,
{{Email:} lindner@ph.tum.de}}
}

\begin{document}

\begin{abstract}
Different future long baseline neutrino oscillation setups are
discussed and the remarkable potential for very precise measurements 
of mass splittings and mixing angles is shown. Furthermore it will 
be possible to make precise tests of MSW effects, which allow to 
determine the sign of $\Delta m^2$. Finally strong limits or 
measurements of leptonic CP violation will be possible.
\vspace{1pc}
\end{abstract}

\maketitle

\renewcommand{\thefootnote}{\fnsymbol{footnote}}
\setcounter{footnote}{1}

\footnotetext{To appear in the proceedings
of the XXth International Conference on Neutrino Physics and Astrophysics, 
``Neutrino 2002'', Munich, Germany, May 25-30, 2002}


\renewcommand{\thefootnote}{\arabic{footnote}}
\setcounter{footnote}{0}


\section{Introduction}
\label{sec:intro}

The existing evidence for atmospheric neutrino oscillations 
includes some sensitivity to the characteristic $L/E$ 
dependence of oscillations \cite{Toshito:2001dk}, and there 
is no doubt that the observed flavour transitions are due 
to neutrino oscillations. Solar neutrinos were also shown to 
undergo flavour transitions \cite{Ahmad:2002jz,Ahmad:2002ka}.
This solves the long standing solar neutrino problem, even 
though the characteristic $L/E$ dependence of oscillation is 
in this case not yet established. However, oscillation is 
under all alternatives by far the most plausible explanation 
and global oscillation fits clearly favour 
the so-called LMA solution for the mass splittings and mixings
\cite{Barger:2002iv,Bandyopadhyay:2002xj,Bahcall:2002hv,%
deHolanda:2002pp}. The CHOOZ reactor experiment \cite{Apollonio:1999ae} 
provides moreover currently the most stringent upper bound for the 
sub-leading $U_{e3}$ element of the neutrino mixing matrix. 
The global pattern of neutrino oscillation parameters seems 
therefore quite well known and one may ask how precise future 
experiments will ultimately be able to measure mass splittings 
and mixings and what can be learned from such precise measurements.

The characteristic length scale $L$ of oscillations is given 
by $\dm{}L/E_\nu=\pi/2$. The atmospheric $\dm{31}$-value leads  
thus to an oscillation length scale $L_{atm}$ as a function of 
energy. For $\dm{31} \simeq 3\cdot 10^{-3}$~eV, and for neutrino 
energies of $E_\nu\simeq 10$~GeV, one finds 
$L_{atm}\simeq {\cal O}(2000)$~km, \ie\ distances and energies 
which can be realized by sending neutrino beams from one point 
on the Earth to another. The solar $\dm{21}$ is for the 
favoured LMA solution about two orders of magnitude smaller than 
the atmospheric $\dm{31}$, resulting for the same energies in 
oscillations at scales $L_{sol} \simeq (10-1000) \cdot L_{atm}$. 
Solar oscillations will thus not fully develop in LBL experiments 
on Earth, but sub-leading effects play nevertheless an important 
role in precision experiments. Another modification comes from 
the fact that the neutrino beams of LBL experiments traverse the 
matter of the Earth. Coherent forward scattering in matter must 
therefore to be taken into account, which makes the analysis 
more involved, but it offers also unique opportunities.

The existing K2K experiment as well as MINOS and CNGS, which are 
both under construction, are a promising first generation of 
LBL experiments. We will discuss here (based on \cite{Huber:2002mx}) 
the remarkable potential byond this first generation. One 
important point is that the increased precision will allow to 
test in detail the three-flavouredness of oscillations. We will 
also see that it will be possible to limit or measure $\theta_{13}$ 
drastically better than today. Next it will be possible to study in 
detail MSW matter effects and to extract in this way $sign(\dm{31})$, 
\ie\ the mass ordering of the neutrino states. For the now favoured 
LMA solution it will be possible to measure leptonic CP violation 
\cite{Dick:1999ed}. The precise neutrino masses, mixings and CP phases 
which can be obtained in this way are extremely valuable information 
about flavour physics, since unlike for quarks these parameters are 
not obscured by hadronic uncertainties. These parameters can be 
evolved with the renormalization group \eg\ to the GUT scale, 
where the rather precisely known parameters can be compared with mass 
models neutrino masses and mixings. Leptonic CP violation is moreover 
related to leptogenesis, the currently most plausible mechanism for 
the generation of the baryon asymmetry of the universe. LBL experiments 
offer therefore in a unique way access to precise knowledge on extremely 
interesting and valuable physics parameters.


\section{Beams and Detectors}
\label{sec:sod} 

LBL experiments have the advantage that both source and  
detector can be kept under precise conditions. 
This includes amongst others for the source a precise knowledge 
of the mean neutrino energy $E_\nu$, the neutrino flux and 
spectrum, as well as the flavour composition and contamination 
of the beam. Another important aspect is whether neutrino and 
anti-neutrino data can be obtained symmetrically such that 
systematical uncertainties cancel in an analysis. 
Precise measurements require also a sufficient luminosity and a 
detector such that enough statistics can be 
obtained. On the detector side one must include further issues, 
like the detection threshold function, energy calibration, resolution, 
particle identification capabilities (flavour, charge, event 
reconstruction, understanding backgrounds). Another source of 
uncertainty in the detection process is the knowledge of neutrino 
cross-sections, especially at low energies \cite{Paschos:2002mb}. 
Source and detector combinations of a future LBL experiment are 
furthermore constraint by the available technology and one 
should keep potential improvements of source and detector 
developments in mind. An example is given by liquid argon detectors 
like ICARUS \cite{Arneodo:jt} which are not included in this study, 
but might become extremely valuable detectors or detector components 
in the future, improving the potential even further.

\begin{figure}
\bc
\includegraphics[width=7cm]{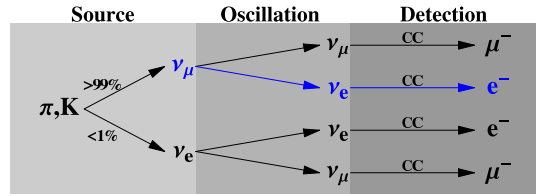} 
\caption{Neutrino production, oscillation and detection via 
charged current interactions for conventional beams and 
superbeams. Most interesting are the $\ruu$ disappearance 
and $\rue$ appearance channels. The $\nu_e$ beam contamination 
at the level of $<1\%$ limits the ability to determine the 
$\rue$ appearance oscillation, since it produces also electrons. 
The $\nu_\tau$ channel is not shown here, but it would become 
very important if tau lepton detection were feasible.}
\label{fig:sbsignal}  
\ec
\end{figure}

The first type of considered sources are conventional neutrino and 
anti-neutrino beams. An intense proton beam is typically directed 
onto a massive target producing mostly pions and some K mesons, 
which are captured by an optical system of magnets in order to obtain 
a beam. The pions (and K mesons) decay in a decay pipe, yielding 
essentially a muon neutrino beam which can undergo oscillations as 
shown in fig.~\ref{fig:sbsignal}. Most interesting are the $\ruu$ 
disappearance channel and the $\rue$ appearance channels.
The neutrino beam is, however, contaminated by approximately
$0.5\%$ electron neutrinos, which also produce electron reactions 
in the disappearance channel, limiting thus the precision in the 
extraction of $\rue$ oscillation parameters. 
The energy spectrum of the muon beam can be controlled over 
a wide range: it depends on the incident proton energy, the 
optical system, and the precise direction of the beam axis 
compared to the direction of the detector. It is possible to 
produce broad band high energy beams, such as the CNGS 
beam~\cite{CNGS1,CNGS2}, or narrow band lower energy beams, 
such as in some configurations of the NuMI beam~\cite{MINOS}.
Reversing the electrical current in the lens system results 
in an anti-neutrino beam. The neutrino and anti-neutrino 
beams have significant differences such that errors do not 
cancel systematically in ratios or differences. The neutrino
and anti-neutrino beams must therefore more or less be considered 
as independent sources with different systematical errors.

So-called ``superbeams'' are based on the same beam dump techniques 
for producing neutrino beams, but at much larger luminosities 
\cite{CNGS1,CNGS2,MINOS,Nakamura:2000uu}. Superbeams are thus a
technological extrapolation of conventional beams, but use a 
proton beam intensity close to the mechanical stability limit of 
the target at a typical thermal power of $0.7\,\mathrm{MW}$ to 
$4\,\mathrm{MW}$. The much higher neutrino luminosity allows the 
use of the decay kinematics of pions to produce so--called 
``off--axis beams'', where the detector is located some degrees 
off the beam axis. This reduces the neutrino flux and the 
average neutrino energy, but leads to a more mono-energetic beam 
and a significant suppression of the electron neutrino contamination. 
Several off--axis superbeams with energies of about $1\,\mathrm{GeV}$ 
to $2\, \mathrm{GeV}$ have been proposed in 
Japan~\cite{Itow:2001ee,Aoki:2002ks},
America~\cite{Para:2001cu}, and Europe~\cite{Gomez-Cadenas:2001eu,Dydak}.

The most sensitive neutrino oscillation channel for sub-leading 
oscillation parameters is the $\rue$ appearance transition. Therefore 
the detector should have excellent electron and muon charged current 
identification capabilities. In addition, an efficient rejection of 
neutral current events is required, because the neutral current 
interaction mode is flavor blind. With low statistics, the magnitude 
of the contamination itself limits the sensitivity to the $\rue$
transition severely, while the insufficient knowledge of its magnitude 
constrains the sensitivity for high statistics. A near detector 
allows a substantial reduction of the background 
uncertainties~\cite{Itow:2001ee,Szleper:2001nj} and plays a 
crucial role in controlling other systematical errors, such as the flux 
normalization, the spectral shape of the beam, and the neutrino cross 
section at low energies. At energies of about $1\,\mathrm{GeV}$, the 
dominant charge current interaction mode is quasi--elastic scattering, 
which suggests that water Cherenkov detectors are the optimal type 
of detector. At these energies, a baseline of about 
$300\,\mathrm{km}$ would be optimal to measure at the first
maximum of the oscillation. At about $2\,\mathrm{GeV}$, there is 
already a considerable contribution of inelastic scattering to the 
charged current interactions, which means that it would be useful to 
measure the energy of the hadronic part of the cross section. This 
favors low--Z hadron calorimeters, which also have a factor of ten 
better neutral current rejection capability compared to water 
Cherenkov detectors~\cite{Para:2001cu}. In this case, the optimum 
baseline is around $600\,\mathrm{km}$. The matter effects are expected
to be small for these experiments for two reasons. First of all, an
energy of about $1\,\mathrm{GeV}$ to $2\,\mathrm{GeV}$ is small 
compared to the MSW resonance energy of approximately $13\,\mathrm{GeV}$ 
in the upper mantle of the Earth. The second reason is that the baseline 
is too short to produce significant matter effects.

The second type of beam considered are so-called neutrino factories,
where muons are stored in the long straight sections of a storage ring.
The decaying muons produce muon and electron anti-neutrinos 
in equal numbers~\cite{Geer:1998iz}. The muons are produced by pion 
decays, where the pions are produced by the same technique as for 
superbeams. After being collected, they have to be cooled and 
reaccelerated very quickly. This has not yet been demonstrated and 
it is a major technological challenge. The spectrum and 
flavor content of the beam are completely characterized by the muon 
decay and are therefore very precisely known~\cite{PDG}. The only 
adjustable parameter is the muon energy $E_\mu$, which is usually 
considered in the range from $20$ to $50\,\mathrm{GeV}$. In a neutrino 
factory it would be possible to produce and store anti-muons in order to 
obtain a CP conjugated beam. The symmetric operation of both beams leads 
to the cancellation or significant reduction of errors and systematical 
uncertainties. We will discuss in the following the neutrino beam, 
which implies always implicitly -- unless otherwise stated -- the 
CP conjugate channel.

\begin{figure}
\bc
\includegraphics[width=7cm]{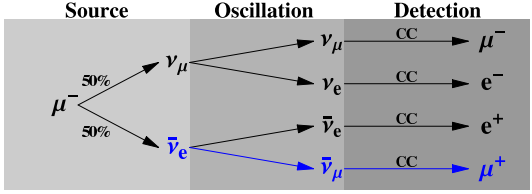} 
\caption{Neutrino production, oscillation and detection via 
charged current interactions for a neutrino factory for one
polarity. $\bar\nu_e$ and $\nu_\mu$ are produced in equal numbers
from $\mu$-decays and can undergo different oscillations. 
The $\ruu$ and $\reub$ channels are most interesting for 
detectors with $\mu$ identification. Note, however, that 
excellent charge identification capabilities are required to
separate ``wrong sign muons'' and ``right sign muons''.
The $\nu_\tau$ oscillation channel is not shown here, but
it would become important for detectors with tau identification
capabilities.}
\label{fig:nufactsignal}  
\ec
\end{figure}

The decay of the muons and the relevant oscillation channels are shown 
in fig.~\ref{fig:nufactsignal}. 
Amongst all flavors and interaction types, 
muon charged current events are the easiest to detect. The appearance channel 
with the best sensitivity is thus the $\reub$ transition, which produces 
so called ``wrong sign muons''. Therefore, a detector must be able to very 
reliably identify the charge of a muon in order to distinguish wrong sign 
muons in the appearance channel from the higher rate of same sign muons in 
the disappearance channels. The dominant charge current interaction in the 
multi--GeV range is deep--inelastic scattering, making a good energy 
resolution for the hadronic energy deposition necessary. Magnetized iron 
calorimeters are thus the favored choice for neutrino factory detectors. 
In order to achieve the required muon charge separation, it is necessary 
to impose a minimum muon energy cut at 
approximately $4\,\mathrm{GeV}$~\cite{Blondel:2000gj,Albright:2000xi}. 
This leads to a significant loss of neutrino events in the range of about 
$4\,\mathrm{GeV}$ to $20\,\mathrm{GeV}$, which means that a high muon 
energy of $E_\mu = 50\,\mathrm{GeV}$ is desirable. The first oscillation 
maximum lies then at approximately $3\,000\,\mathrm{km}$. Matter effects 
are sizable at this baseline and energy, and the limited knowledge of the 
Earth's matter density profile becomes an additional source of errors.


\section{The Oscillation Framework}
\label{sec:osc}

The general three neutrino oscillation probabilities in matter
are quite lengthy expressions. It is useful to simplify them
such that an analytic understanding becomes possible.
Quantitative results should, however, be obtained in numerical 
calculations using the full expressions. The key to an analytic 
simplification is to expand the oscillation probabilities 
in small quantities, namely $\alpha=\dm{21}/\dm{31} \simeq 10^{-2}$ 
and $\sin^2 2\theta_{13} \leq 0.1$. The matter effects can be 
parametrized by the dimensionless quantity
$\hat A=A/\dm{31}=2VE/\dm{31}$, where $V=\sqrt{2}G_F n_e$.
Using $\Delta \equiv \Delta_{31}$, the leading terms in this 
expansion are \cite{Cervera:2000kp,Freund:2001pn,Freund:2001ui}

\begin{eqnarray}
P(\nu_\mu \rightarrow \nu_\mu) \approx \hspace*{4.2cm}\   \nonumber\\
\hspace*{5mm}
 1 - c^2_{13} {\hat s}^2_{23} \sin^2 {\Delta}  
+ 2~{\alpha}~  c^2_{13} c^2_{12} {\hat s}^2_{23} {\Delta} \cos{\Delta}
\label{eq:Pdis}
\end{eqnarray}
\begin{eqnarray}
P(\nu_e \rightarrow \nu_\mu) \approx {\hat s}^2_{13} s^2_{23}
~{\scriptstyle
\frac{\sin^2({(1\!\!-\!\hat{A})}{\Delta})}
{{(1\!\!-\!\hat{A})^2}} }\hspace*{1.7cm}  \nonumber\\
\hspace*{5mm} 
\pm~ s_\delta {{\hat s}_{13}}~{\alpha}~ {\hat s}_{12} c_{13} {\hat s}_{23}~
{\scriptstyle
\sin({\Delta})
\frac{
\sin({\hat{A}}{\Delta})\sin({(1\!\!-\!\hat{A})}{\Delta})}
{{\hat{A}(1\!\!-\!\hat{A})}}}
\nonumber \\
\hspace*{5mm}
+~ c_\delta {\hat s}_{13}~ {\alpha}~ {\hat s}_{12} c_{13}  
{\hat s}_{23}~
{\scriptstyle
\cos({\Delta})\frac{\sin({
\hat{A}}{\Delta})\sin({(1\!\!-\!\hat{A})}{\Delta})}{{\hat{A}(1\!\!-\!\hat{A})}}
}
\nonumber \\
+~  {\alpha^2}~ {\hat s}^2_{12} c^2_{23} 
\frac{\sin^2({\hat{A}}{\Delta})}{{\hat{A}^2}} \hspace*{2.85cm}
\label{eq:Pap}
\end{eqnarray}
where $s_{ij} = \sin\theta_{ij}$, $c_{ij} = \cos\theta_{ij}$,
${\hat s}_{ij} = \sin 2\theta_{ij}$, ${\hat c}_{ij} = \cos 2\theta_{ij}$,
$s_\delta=\sin_\delta$, $c_\delta=\cos_\delta$,
and where in eq.~(\ref{eq:Pap}) ``$+$'' stands for neutrinos and ``$-$'' 
for anti-neutrinos.
The most important feature of eq.~(\ref{eq:Pap}) is that all interesting 
effects in the $\reu$ transition depend crucially on $\theta_{13}$. The 
size of $\sin^2 2\theta_{13}$ determines thus if the total transition rate, 
matter effects, effects due to the sign of $\dm{31}$ and CP violating
effects are measurable. One of the most important questions for future 
LBL experiments is therefore how far experiments can push the 
$\theta_{13}$ limit below the current CHOOZ bound of approximately  
$\sin^2 2\theta_{13}<0.1$.


Before we discuss further important features of eqs.~(\ref{eq:Pdis}) 
and (\ref{eq:Pap}) in more detail we would like to comment once more
on the underlying assumptions and the reliability of these equations.
First eqs.~(\ref{eq:Pdis}) and (\ref{eq:Pap}) are an expansion in 
terms of the small quantities $\alpha$ and $\sin 2\theta_{13}$.
Higher order terms are suppressed at least by another power of 
one of these small parameters and these corrections are thus typically
at the percent level. The matter corrections in eqs.~(\ref{eq:Pdis}) 
and (\ref{eq:Pap}) are derived for constant average matter density.
Numerical test have shown that this approximation works quite well
as long as the matter profile is reasonably smooth. A number of
very interesting effects existing in general non-constant matter 
distributions are therefore only small theoretical uncertainties. 
An example is given by asymmetric matter profiles, which lead to 
interesting T-violating effects \cite{Akhmedov:2001kd}, but this 
does not play a role here since the Earth is sufficiently symmetric. 

Note that all results which will be shown later are based on 
numerical simulations of the full problem in matter. These results 
do therefore not depend on any approximation. Eqs.~(\ref{eq:Pdis}) 
and (\ref{eq:Pap}) will only be used to understand the problem 
analytically, which is extremely helpful in order to oversee the 
six (or more) dimensional parameter space. The full numerical 
analysis and eqs.~(\ref{eq:Pdis}) and (\ref{eq:Pap}) depend,
however, on the assumption of a standard three neutrino scenario. 
It is thus assumed that the LSND signal \cite{Church:2002tc}
will not be confirmed by the MiniBooNE experiment \cite{Hawker:pt}.
Neutrinos could in principle decay, which would make the analysis 
much more involved. It is assumed in this article that neutrinos 
are stable, and a combined treatment of oscillation and decay 
\cite{Lindner:2001fx} would be much more involved. Neutrinos might 
further have unusual properties and might, for example, violate CPT. 
In that case neutrinos and anti-neutrinos could have different 
properties and LBL experiments can give very interesting limits 
on this possibility \cite{Bilenky:2001ka}, but we will assume in 
this study that CPT is preserved.


\section{Correlations and Degeneracies}
\label{sec:CD}

Eqs.~(\ref{eq:Pdis}) and (\ref{eq:Pap}) exhibit certain parameter 
correlations and degeneracies, which play an important role in the 
analysis of LBL experiments, and which would be hard to understand
in a purely numerical analysis of the high dimensional parameter 
space. The most important properties are:

\begin{itemize}
\item
First we observe that eqs.~(\ref{eq:Pdis}) and (\ref{eq:Pap})
depend only on the product $\alpha\cdot \sin 2\theta_{12}$ or 
equivalently $\dm{21}\cdot \sin 2\theta_{12}$. This are the 
parameters related to solar oscillations which will be taken
as external input. The fact that only the product enters, implies
that it may be better determined than the product of the 
measurements of $\dm{21}$ and $\sin 2\theta_{12}$. 
\item
Next we observe in eq.~(\ref{eq:Pap}) that the second and third 
term contain both a factor $\sin(\hat A\Delta)$, while the last 
term contains a factor $\sin^2(\hat A\Delta)$. Since 
$\hat A\Delta= 2VL$, we find that these factors depend only on 
$L$, resulting in a ``magic baseline'' when $2VL_{magic}=\pi/4V$, 
where $\sin(\hat A\Delta)$ vanishes. At this magic baseline only 
the first term in eq.~(\ref{eq:Pap}) survives and 
$P(\nu_e \rightarrow \nu_\mu)$ does no longer depend on $\delta$, 
$\alpha$ and $\sin 2\theta_{12}$. This is in principle very 
important, since it implies that $\sin^2 2\theta_{13}$ can be 
determined at the magic baseline from the first term of 
eq.~(\ref{eq:Pap}) whatever the values and errors of $\delta$, 
$\alpha$ and $\sin 2\theta_{12}$ are. For the matter density of 
the Earth we find 
\beq
L_{magic}= \pi/4V \simeq 8100~\mathrm {\rm km}~,
\label{eq:Lmagic}
\eeq
which is an amazing number, since the value of $V$ could be such 
that $L_{magic}$ is very different from the scales under
discussion.
\item
Next we observe that only the second and third term of 
eq.~(\ref{eq:Pap}) depend on the CP phase $\delta$, and 
both terms contain a factor $\sin 2\theta_{13}\cdot\alpha$, 
while the first and fourth term of eq.~(\ref{eq:Pap})
do not depend on the CP phase $\delta$ and contain factors of  
$\sin^2 2\theta_{13}$ and $\alpha^2$, respectively.
The extraction of CP violation is thus always suppressed by 
the product $\sin 2\theta_{13}\cdot\alpha$ and the CP violating 
terms are furthermore obscured by large CP independent terms if 
either $\sin^2 2\theta_{13} \ll \alpha^2$ or
$\sin^2 2\theta_{13} \gg \alpha^2$. The determination of the 
CP phase $\delta$ is thus best possible if 
$\sin^2 2\theta_{13} \simeq 4\theta^2_{13} \simeq \alpha^2$.
\item
The last term in eq.~(\ref{eq:Pap}),
which is proportional to $\alpha^2=(\dm{21})^2/(\dm{31})^2$,
dominates for tiny $\sin^2 2\theta_{13}$. The error 
of $\dm{21}$ limits therefore for small $\sin^2 2\theta_{13}$ the 
parameter extraction.
\item
Eqs.~(\ref{eq:Pdis}) and (\ref{eq:Pap}) suggest the existence of 
degeneracies, \ie\ for given $L/E$ parameter sets with identical 
oscillation probabilities. An example is given by a simultaneous 
replacement of neutrinos by anti-neutrinos and $\dm{31}\rightarrow -\dm{31}$. 
This is equivalent to changing the sign of the second term of 
eq.~(\ref{eq:Pap}) and replacing $\alpha \rightarrow -\alpha$ and
$\Delta \rightarrow -\Delta$, while $\hat A \rightarrow \hat A$.
Eqs.~(\ref{eq:Pdis}) and (\ref{eq:Pap}) are unchanged, but this does
not constitute a degeneracy, since neutrinos and anti-neutrinos can 
be distinguished experimentally. 
\item
The first real degeneracy \cite{Barger:2001yr} can be seen in the 
disappearance probability eq.~(\ref{eq:Pdis}), which is invariant 
under $\theta_{23}\rightarrow \pi/2 - \theta_{23}$.
Note that the second and third term in eq.~(\ref{eq:Pap}) are not 
invariant under this transformation, but this change in the sub-leading 
appearance probability can approximately be compensated by small 
parameter shifts. However, the degeneracy can in principle be lifted
with precision measurements in the disappearance channels.
\item
The second degeneracy can be found in the appearance probability 
eq.~(\ref{eq:Pap}) in the ($\delta-\theta_{13}$)-plane 
\cite{Burguet-Castell:2001ez}. In terms of $\theta_{13}$ (which is 
small) and $\delta$ the four terms of eq.~(\ref{eq:Pap}) have the 
structure  
\begin{eqnarray}
P(\nu_e \rightarrow \nu_\mu) \approx 
\theta^2_{13}\cdot F_1 \hspace*{2.0cm}\nonumber \\ 
\hspace*{5mm}+ \theta_{13}\cdot (\pm~\sin\delta F_2 + \cos\delta F_3) + F_4
\label{eq:degth13}
\end{eqnarray}
where the quantities $F_i$, $i=1,..,4$ contain all the other parameters. 
The requirement $P(\nu_e \rightarrow \nu_\mu)=const.$ leads for both 
neutrinos and anti-neutrinos to parameter manifolds of degenerate
or correlated solutions. Having both neutrino and anti-neutrino beams, 
the two channels can be used independently, which is equivalent to 
considering simultaneously eq.~(\ref{eq:degth13}) for $F_2\equiv 0$ 
and $F_3\equiv 0$. The requirement that these probabilities are now 
independently constant, \ie\ $P(\nu_e \rightarrow \nu_\mu)=const.$ 
for $F_2\equiv 0$ and $F_3\equiv 0$, leads to more constraint manifolds 
in the ($\delta-\theta_{13}$)-plane, but some degeneracies still survive.
\item
The third degeneracy \cite{Minakata:2001qm} is given by the fact 
that a change in sign of $\dm{31}$ can essentially be compensated 
by an offset in $\delta$. Therefore we note again that the transformation 
$\dm{31} \rightarrow -\dm{31}$ leads to $\alpha \rightarrow -\alpha$,
$\Delta \rightarrow -\Delta$ and $\hat A \rightarrow -\hat A$. 
All terms of the disappearance probability, eq.~(\ref{eq:Pdis}), are 
invariant under this transformation. The first and fourth term in 
the appearance probability eq.~(\ref{eq:Pdis}), which do not depend
on the CP phase $\delta$, are also invariant. The second and third 
term of eq.~(\ref{eq:Pdis}) depend on the CP phase and change by the 
transformation $\dm{31} \rightarrow -\dm{31}$. The fact that these 
changes can be compensated by an offset in the CP phase $\delta$ 
is the third degeneracy.
\item
Altogether there exists an eight-fold degeneracy 
\cite{Barger:2001yr}, as long as only the $\ruu$, $\ruub$,
$\reu$ and $\reub$ channels and one fixed $L/E$ are considered. 
However, the structure of eqs.~(\ref{eq:Pdis}) and (\ref{eq:Pap}) 
makes clear that the degeneracies can be broken by using in a 
suitable way information from different $L/E$ values. This can be 
achieved in total event rates by changing $L$ or $E$ 
\cite{Burguet-Castell:2002qx,Barger:2002rr}, but it can in principle 
also be done by using information in the event rate spectrum of a
single baseline $L$, which requires detectors with very good energy 
resolution \cite{Freund:2001ui}. Another way to break the 
degeneracies is to include further oscillation channels in the 
analysis (``silver channels'') \cite{Donini:2002rm,Burguet-Castell:2002qx}.
\end{itemize}
The discussion shows the strength of the analytic approximations, 
which allow to understand the complicated parameter interdependence. 
It also helps to optimally plan experimental setups and to find 
strategies to resolve the degeneracies.


\section{Event Rates}
\label{sec:evrates} 

The experimentally observed events must be compared with the 
theoretical expressions, which depend only indirectly on the above 
oscillation probabilities \cite{Huber:2002mx}. Every event can be 
classified by the information on the flavor of the detected neutrino 
and the type of interaction. The particles detected in an experiment 
are produced by neutral current (NC), inelastic charged current (CC) 
or quasi--elastic charged current (QE) interactions. The contribution 
to each mode depends on a number of factors, like detector type, the 
neutrino energy and flavour. To determine realistic event
rates we compute first for each neutrino flavor and energy bin
the number of events for each type of interaction in the fiducial 
mass of an ideal detector. Next the deficiencies of a real detector
are included, like limited event reconstruction capabilities. 
The combined description leads to the differential event rate 
spectrum for each flavor and interaction mode as it would be seen 
by a detector which is able to separate all these channels.
Finally different channels must be combined, if they can not
be observed separately. This can be due to physics, \eg, due to
the flavor--blindness of NC interactions, or it can be a consequence
of detector properties, \eg, due to charge misidentification.
In order to include backgrounds, the channels are grouped in 
an experiment specific way into pairs of signal and background. 
The considered backgrounds are NC--events which are misidentified 
as CC--events and CC--events identified with the wrong flavor or 
charge. For superbeams the background of CC--events coming from 
an intrinsic contamination of the beam is included. 
Finally all available signal channels are combined in a global 
analysis in order to optimally extract the physics parameters.
The relevant channels are for a neutrino factory for each polarity 
of the beam the $\nu_\mu$--CC channel (disappearance) and 
$\bar{\nu}_\mu$--CC channel (appearance) event rate spectra. The 
backgrounds for these signals are NC events for all flavours and 
misidentified $\nu_\mu$--CC events. For superbeam experiments the 
signal is for each polarity of the beam given by the $\nu_\mu$--QE 
channel (disappearance) and ${\nu}_e$--CC channel (appearance). 
The backgrounds are here NC events for all flavors, misidentified 
$\nu_\mu$--CC events, and, for the  ${\nu}_e$--CC channel, the 
$\nu_e$--CC beam contamination.


\section{The Considered LBL Setups}
\label{sec:LBLpot}

The discussed sources and detectors allow different LBL experiments 
and it is interesting to compare their physics potential on an 
equal and as realistic as possible footing. Studies at
the level of probabilities are not sufficient and the true potential 
must be evaluated at the level of event rates as described in 
section~\ref{sec:evrates}, with realistic assumptions about the 
beams, detectors and backgrounds. We present now the results of such 
an analysis which is essentially based on reference \cite{Huber:2002mx}, 
where we calculate the oscillation probabilities with 
the exact three neutrino oscillation formulae in matter numerically, 
\ie\ we use the approximations for the probabilities in eqs.~(\ref{eq:Pdis}) 
and (\ref{eq:Pap}) only for a qualitative understanding. All results 
shown below are therefore not affected by approximations which were 
made in the derivation of the approximate analytic oscillation formulae 
eqs.~(\ref{eq:Pdis}) and (\ref{eq:Pap}).
Sensitivities etc. are defined by the ability to re-extract the 
physics parameters from a simulation of event rates. Therefore
event rate distributions are generated for any possible parameter 
set. Subsequently a combined fit to these event rate distributions 
is performed simultaneously for the appearance and disappearance 
channels for both polarities. This procedure uses all the available 
information in an optimal way. It includes spectral distributions when 
present, and it reduces to a fit of total rates when the total event 
rates are small. Adequate statistical methods as described in 
\cite{Huber:2002mx} are used in order to deal with event distributions 
which have in some regions small event rates per bin. Systematical 
uncertainties are parametrized and external input from geophysics is used 
in form of the detailed matter profile and its errors, which are included 
in the analysis \cite{Huber:2002mx}. The ability to re-extract in a 
simulation of the full experiment the input parameters which were 
used to generate event rate distributions is used define sensitivities 
and precision.

It is important to include in the analysis external 
information. The discussed LBL experiments could in principle measure 
the solar $\dm{21}$ and mixing angle $\theta_{12}$. However, the precision 
which can be obtained can not compete with the expected measurement of 
KamLand \cite{Barger:2000hy}. We include therefore as external input 
the assumption that KamLand measures the solar parameters in the center 
of the LMA region with typical errors. Otherwise all unknown parameters 
(like the CP phase) will not be constrained and are therefore left free, 
with all parameter degeneracies and error correlations taken into account. 
All nuisance parameters are integrated out and a projection on the 
parameter of interest is performed. Altogether we are dealing with 
six free parameters.

We include the beam characteristics of the three considered sources 
as well as uncertainties of these beam parameters, \ie\ for the
conventional JHF and NuMI off-axis beams uncertainties in the 
$\nu_e$--background and for all beams flux uncertainties 
\cite{Itow:2001ee,Para:2001cu,Geer:1998iz}.
As detectors we consider water Cherenkov detectors, low-Z 
calorimeters and magnetized iron detectors. 
For magnetized iron calorimeters it is
important to include realistic threshold effects. We use a linear 
rise of the efficiency between $4\,\mathrm{GeV}$ and $20\,\mathrm{GeV}$
and we study the sensitivity to the threshold position. 
We do not include liquid Argon TPCs in our analysis, but they would 
certainly be an important detector if this technology will work.
The considered beams and detectors allow now different interesting
combinations which are listed in table~\ref{tab:scenarios}. JHF-SK
is the planned combination of the existing SuperKamiokande detector
and the JHF beam, while JHF-HK is the combination of an upgraded 
JHF beam with the proposed HyperKamiokande detector. With typical
parameters, JHF-HK is altogether about $95$ times more integrated 
luminosity than JHF-SK, and we assume that it operates partly with 
the anti-neutrino beam. Water Cherenkov detectors are ideal for the
JHF beam, since charged current quasi elastic scattering is dominating.
A low-Z calorimeter is proposed for the NuMI off-axis beam,
which is better here, since the energy is higher
and there is already a considerable contribution of inelastic 
charged current interactions. 
NuFact-I is an initial neutrino factory, while NuFact-II is a fully 
developed machine, with $42$ times the luminosity of NuFact-I 
\cite{Itow:2001ee,Para:2001cu,Blondel:2000gj}. Deep inelastic scattering
dominates for these even higher energies and magnetized iron 
detectors are therefore considered in combination with neutrino factories.

\begin{table}
\bc
\begin{tabular}{|l|c|c|c|}
\hline
acronym & detector & $L$ & $L/E_{\mathrm{peak}}$\\
\hline\hline
{\bf JHF-SK} & water Cherenkov & 295 & 378 \\ 
\hline
{\bf NuMI} & low-Z & 735 & 337 \\ 
\hline
{\bf NuFact-I} & $10$~kt mag. iron & 3000 & 90 \\ 
\hline
{\bf JHF-HK} & water Cherenkov & 735 & 295 \\ 
\hline
{\bf NuFact-II} & $40$~kt mag. iron & 3000 & 90 \\
\hline
\end{tabular}
\ec
\caption{The considered combinations of beams and detectors 
and their acronyms.}
\label{tab:scenarios}
\end{table}

\section{Results}
\label{sec:results}

A realistic analysis of future LBL experiments requires a number 
of different aspects to be taken into account. It should be clear 
from the discussion above that it is not sufficient to quote limits 
which are based on oscillation probabilities or merely on the 
statistics of a single channel without backgrounds or systematics. 
Depending on the position in the space of physics parameters the 
degeneracies or correlations, the backgrounds, the systematics or 
statistics may be the limiting factor. A reliable comparative 
study of the discussed LBL setups requires therefore a detailed 
analysis of the six dimensional parameter space, which 
includes all these effects on the same footing \cite{Huber:2002mx}.

\begin{figure}[tb]
\bc
\includegraphics[width=7.2cm]{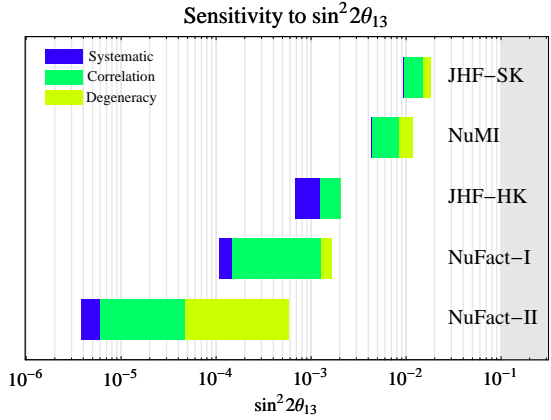} 
\ec
\caption{The $\sin^2 2 \theta_{13}$ sensitivity for all setups 
defined in section~\ref{sec:LBLpot} at the 90\%  confidence level for 
$\dm{31}= 3\cdot 10^{-3} \, \mathrm{eV}^2$ and $\sin^2 2 \theta_{23}=0.8$.
The plot shows the deterioration of the sensitivity limits as the
different error sources are successively switched on. The left
edge of the bars is the sensitivity statistical limit. This 
limit gets reduced as systematical, correlational and degeneracy
errors are switched on. The right edge is the final sensitivity 
limit \cite{Huber:2002mx}.}
\label{fig:th13exec}
\end{figure}

There is excellent precision for the leading oscillation parameters
$\dm{31}$ and $\sin^2 2\theta_{23}$, which will not be further discussed 
here. The more interesting sensitivity to the sub-leading parameter 
$\sin^2 2\theta_{13}$ depends on what will be found for $\dm{31}$ 
and $\dm{21}$. Assuming that the leading parameters are measured to be 
$\dm{31}= 3\cdot 10^{-3}~\mathrm{eV}^2$, $\sin^2 2 \theta_{23}=0.8$
and that KamLand measures the solar parameters at the current best 
fit point of the LMA region, \ie\ $\dm{21}= 6\cdot 10^{-5}~\mathrm{eV}^2$ 
and $\sin 2\theta_{12} = 0.91$, we can make a comparison of the 
$\sin^2 2\theta_{13}$ sensitivity limit for the different setups. The 
result is shown in fig.~\ref{fig:th13exec}. The individual contributions
of different sources of uncertainties are shown for every experiment
and the left edge of every band in fig.~\ref{fig:th13exec} corresponds 
to the sensitivity limit which would be obtained purely on statistical 
grounds. This limit is successively reduced by adding the systematical 
uncertainties of each experiment, the correlational errors and finally 
the degeneracy errors. The right edge of each band constitutes the 
final error for the experiment under consideration.
It is interesting to see how the errors of the different setups are
composed. There are different sensitivity reductions due to systematical
errors, correlations and degeneracies.
The largest sensitivity loss due to correlations and degeneracies 
occurs for NuFact-II, which is mostly a consequence of the 
uncertainty of $\dm{21}$, which translates into an $\alpha$ uncertainty,
and which dominates the appearance probability eq.~(\ref{eq:Pap}) for 
small $\sin^22\theta_{13}$. Note that it is in principle
possible to combine different experiments. If done correctly, this 
allows to eliminate part or all of the correlational and degeneracy 
errors \cite{Burguet-Castell:2002qx}.

\begin{figure}[tb]
\bc
\includegraphics[width=5.1cm,angle=-90]{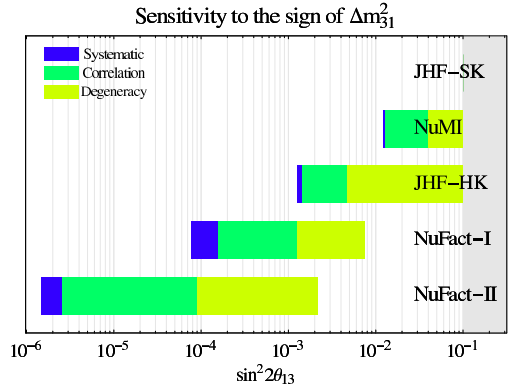} 
\ec
\caption{The $\sin^2 2\theta_{13}$ sensitivity region to 
$sign(\dm{31})$ for the setups defined in section~\ref{sec:LBLpot}.
The left edge of the bars are the statistical sensitivity limits
which are successively reduced by systematical, correlational and
degeneracy errors. The right edge of the bars is the final limit.}
\label{fig:signex}
\end{figure}

Another challenge of future LBL experiments is to measure $sign(\dm{31})$ 
via matter effects and the sensitivity which can be obtained for the setups 
under discussion is shown in fig.~\ref{fig:signex}.
Taking all correlational and degeneracy errors into account we 
can see that it is very hard to determine $sign(\dm{31})$ with 
the considered superbeam setups. The main problem is the degeneracy 
with $\delta$, which allows always the reversed $sign(\dm{31})$ 
for another CP phase. Note, however, that the situation can in
principle be improved if different superbeam experiments were 
combined such that this degeneracy error could be removed. 
Neutrino factories perform considerably better on $sign(\dm{31})$,
particularly for larger baselines. Combination
strategies would again lead to further improvements.

Coherent forward scattering of neutrinos and the corresponding 
MSW matter effects are so far experimentally untested. It is 
therefore very important to realize that matter effects will not 
only be useful to extract $sign(\dm{31})$, but that they allow 
also detailed tests of coherent forward scattering of neutrinos. 
This has been studied in detail in \cite{Freund:2001ui,Freund:1999gy,%
Freund:2000ti,Freund:vt}.

The Holy Grail of LBL experiments is the measurement of leptonic
CP violation.
\begin{figure}[tb]
\bc
\includegraphics[width=5.1cm,angle=-90]{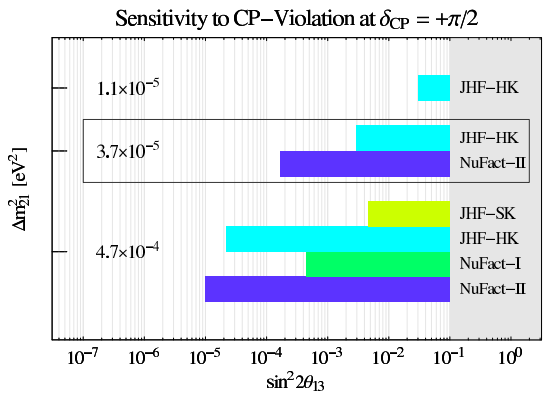} 
\ec
\caption{The $\sin^2 2 \theta_{13}$ sensitivity range for CP violation 
of the considered setups at 90\% confidence level and for different
$\Delta m_{21}^2$ values. The upper row corresponds to the lower bound 
of $\Delta m_{21}^2=1.1\times 10^{-5}~\mathrm{eV}^2$, the bottom row
to the upper bound $\Delta m_{21}^2=4.7\times 10^{-4}~\mathrm{eV}^2$,
and the middle row to the best LMA fit, 
$\Delta m_{21}^2=3.7\times 10^{-5}~\mathrm{eV}^2$. 
Cases which do not have CP sensitivity are omitted from this plot.
The chosen parameters are $\delta=+\pi/2$, 
$\dm{31}= 3\cdot 10^{-3}~\mathrm{eV}^2$, $\sin^2 2 \theta_{23}=0.8$,
and a solar mixing angle corresponding to the current best fit
in the LMA regime \cite{Huber:2002mx}.}
\label{fig:CPexec}
\end{figure}
The $\sin^2 2\theta_{13}$ sensitivity range for measurable CP violation 
is shown in fig.~\ref{fig:CPexec} for $\delta=\pi/2$ for the different
setups and for different values of $\dm{21}$. It can be seen that 
measurements of CP violation are in principle feasible both with high 
luminosity superbeams as well as advanced neutrino factories. However, 
the sensitivity depends in a crucial way on $\dm{21}$. For a low value 
$\dm{21}=1.1~10^{-5}~\mathrm{eV}^2$, the sensitivity is almost completely 
lost, while the situation would be very promising for the largest considered 
value $\dm{21}=4.7~10^{-4}~\mathrm{eV}^2$. For a measurement of leptonic 
CP violation it would therefore be extremely exciting and promising if 
KamLand would find $\dm{21}$ on the high side of the LMA solution (the 
so-called HLMA case). 
The sensitivities shown in fig.~\ref{fig:CPexec} depend on the choice
for $\delta$. The value which was used here was $\delta=\pi/2$ and
and the sensitivities become become worse for small CP phases close to 
zero or $\pi$.

\section{Conclusions}
\label{sec:Conclusio}

Future long baseline neutrino oscillation experiments will lead
to precision neutrino physics. A basic fact which 
makes this possible is that the atmospheric mass splitting
$\dm{31}\simeq\Delta{m^2_\mathrm{atm}}$ leads for typical neutrino 
energies $E_\nu\simeq 1-100$~GeV to oscillation baselines in 
the range $100$~km to $10000$~km. Beams have the advantage, 
that unlike the sun or the atmosphere, they can be controlled 
very precisely. Combined with equally precise detectors and an 
adequate oscillation framework (including three neutrinos and 
matter effects) must be used. There exist different interesting 
sources for long baseline oscillation experiments, like reactors 
or $\beta$-beams, but we restricted the discussion here to superbeams 
and neutrino factories. 
We presented the issues which enter into 
realistic assessments of the potential of such experiments. 
The discussed experiments turned out to be very promising and 
they lead to very precise measurements of the leading oscillation
parameters $\dm{31}$ and $\sin^2 2\theta_{23}$. We discussed
in detail how the different setups lead to very interesting 
measurements or limits $\theta_{13}$ and $\delta$. It will also 
be possible to perform impressive tests of Earth matter effects, 
allowing to extract $sign(\dm{31})$. The discussed setups have
an increasing potential and increasing technological challenges,
but it seems possible to built them in stages. The shown results
are valid for each individual setup and future results should
of course be included in the analysis. This would be especially
important if more LBL experiments were built and depending on
previous results there exist different optimization strategies.
In the short run the expected results from KamLand are extremely
important and will have considerable impact. First it will 
become clear if $\dm{21}$ lies in the LMA regime, which is 
very important since realistically, CP violation can only be 
measured then. Within the LMA solution it is also very important
if $\dm{21}$ lies close to the current best fit, on the high or 
on the low side. A value of $\dm{21}$ on the high side (HLMA) would
be ideal, since it would guarantee an extremely promising LBL
program with a chance to see leptonic CP violation already with 
the JHF beam in the next decade.



\end{document}